\documentclass[aps,floats,floatfix,amssymb,prd,onecolumn,superscriptaddress,nofootinbib,notitlepage,11pt]{revtex4-1}
\usepackage[english]{babel}
\usepackage{latexsym,amssymb,amsmath,amsfonts,physics,url,mathrsfs}
\usepackage{graphicx}
\usepackage{comment}
\usepackage[table]{xcolor}
\usepackage{orcidlink}
\usepackage{tikz}
\usepackage[caption=false,labelformat=empty]{subfig}

\newcommand{\be}{\begin{equation}}
\newcommand{\ee}{\end{equation}}

\hypersetup{
     colorlinks   = true,
     citecolor    = violet,
     urlcolor     = violet,
     linkcolor    = violet}

\newcommand{\papertitle}{Nonlinearities in Kerr Black Hole Ringdown from the Penrose Limit}

\begin{document}

\title[]{\papertitle}

\author{Davide Perrone\orcidlink{0000-0003-4430-4914}}
\affiliation{Department of Theoretical Physics and Gravitational Wave Science Center,  \\
24 quai E. Ansermet, CH-1211 Geneva 4, Switzerland}

\author{Alex Kehagias\orcidlink{ 0000-0001-6080-6215}}
\affiliation{Physics Division, National Technical University of Athens, Athens, 15780, Greece}

\author{Antonio Riotto\orcidlink{0000-0001-6948-0856}}
\affiliation{Department of Theoretical Physics and Gravitational Wave Science Center,  \\
24 quai E. Ansermet, CH-1211 Geneva 4, Switzerland}


\begin{abstract}
\noindent
We provide a fully analytical approach to calculate the nonlinearities of the gravitational waves in the ringdown of a Kerr black hole in the eikonal limit. 
The  corresponding quasi-normal modes are associated to the orbits of a closed circular null geodesic and the problem can be analyzed by taking the Penrose limit around it. 
We calculate analytically the amplitude and the phase of the quadratic quasi-normal modes as well as its dependence on the black hole spin.
\end{abstract}

\maketitle

\section{Introduction}
\noindent
Quasi-Normal Modes (QNMs) are the hallmark of black holes, describing the proper response of these objects to any external perturbation (for a recent review, see Ref. \cite{Berti:2025hly}). Through the study of this response, we can extract various pieces of information about the last phase of the coalescence of two compact objects (ringdown) that culminate with the formation of a black hole.
Characterizing the ringdown phase represents a crucial step of the program to study and understand gravitational waves data from mergers \cite{LIGOScientific:2020tif,LIGOScientific:2021sio,Capano:2021etf, Finch:2022ynt, Isi:2022mhy, Cotesta:2022pci, Siegel:2023lxl} and to prepare the analysis for future ground- and space-based gravitational wave detectors  \cite{Berti:2005ys, Ota:2019bzl, Bhagwat:2019dtm, Pitte:2024zbi}.

Albeit the program has been successful with current data, multiple recent studies \cite{London:2014cma,Mitman:2022qdl, Cheung:2022rbm, Ma:2022wpv, Redondo-Yuste:2023seq, Cheung:2023vki, Zhu:2024rej} highlighted how the nonlinearities in gravity can modify the spectrum of the QNMs that will be observed, in particular by adding additional spectral peaks in correspondence of the linear frequency sum, with sizable amplitudes.

Both numerical and analytical works have been conducted to study the nonlinear modes \cite{Kehagias:2023ctr,Redondo-Yuste:2023seq,Cheung:2023vki,Perrone:2023jzq,Zhu:2024rej,Ma:2024qcv,Bourg:2024jme,Bucciotti:2024zyp,Khera:2024yrk,Kehagias:2024sgh,Bucciotti:2025rxa,bourg2025quadraticquasinormalmodesnull,BenAchour:2024skv,Kehagias:2025xzm,Ling:2025wfv,Kehagias:2025tqi}, focusing mainly on the quadratic quasi-normal mode induced by two $\ell=2$ multipoles, which has a non-negligible possibility of being detected within future third generation detectors \cite{Lagos:2024ekd,Yi_2024,PhysRevD.109.064075,shi2024detectabilityresolvabilityquasinormalmodes}.

Going further and study the higher nonlinear modes is technically and computationally challenging.
Nonetheless it is possible to build some intuition and obtain approximate results by performing a large angular momentum or ``eikonal'' limit, which has been intensively studied for the linear case \cite{PhysRevD.30.295,PhysRev.166.1263,Cardoso_2009,Dolan_2010, Dolan_2018,Hadar:2022xag,Fransen:2023eqj}. 
In this work we build on our previous results \cite{Kehagias:2025ntm} for a Schwarzschild black hole and we extend it to the metric of a Kerr black hole, studying the spin dependence of the nonlinear QNMs amplitude.
We also provide a robust approach with an expansion in large angular momentum, presenting the leading result in this limit.

The paper is organized as follows.
In section \ref{sec:kerr_null_geodesics} we set the notation and review the main properties of rotating black holes described by the Kerr metric. 
In section \ref{sec:penrose_limit} we perform the so-called Penrose limit to write an effective metric in a neighborhood of a closed null geodesic. 
In section \ref{sec:linear_solution} we study the equations of motion of a linear perturbation in the background found from the Penrose limit and match it with the asymptotic solution at $\mathcal{I}^+$. 
In section \ref{sec:noneinear_solution} we  solve the equations of motion for a gravitational perturbation sourced by the linear solution performing the matching with the asymptotic solution also for the nonlinear case. 
We study the resulting nonlinear amplitude ratio between the nonlinear solution and the linear solution squared in section~\ref{sec:amp_phase}. 
We conclude with section \ref{sec:conclusions}.

We will work with geometrical units $c=G=1$ and with a mostly plus signature $(-,+,+,+)$.

\section{Kerr metric and null geodesics}
\label{sec:kerr_null_geodesics}
\noindent
The Kerr metric describes a rotating black hole in the vacuum and can be written in Boyer-Lindquist coordinates as
\begin{align}
    \dd s^2 =& - \left(1- \frac{2M r}{\rho^2}\right)\dd t^2 - \left(\frac{4 a M r\sin^2\theta}{\rho^2}\right)\dd t \dd \phi + \frac{\rho^2}{\Delta}\dd r^2+  \nonumber\\
    &\rho^2 \dd \theta^2 +\sin^2\theta \left(r^2+a^2 + \frac{2 M a^2 r\sin^2 \theta}{\rho^2}\right)\dd\phi^2,
\end{align}
where $M$ is the mass of the black hole, $aM$ is its angular momentum, and
\begin{equation}
    \rho^2=r^2+a^2 \cos^2\theta,\quad \Delta=r^2 -2Mr+a^2.
\end{equation}
Before the calculation of the Penrose limit we need to study the null geodesics on the Kerr black hole background.

The equations of motion for null rays can be written as \cite{Dolan_2010, Chandrasekhar:1985kt, Teo_2021, Teo:2003ltt, Misner:1973prb}
\begin{align}
    \rho^2 \dot t &= \frac{\Sigma^2 E - 2M a r L_z}{\Delta},\nonumber\\
    \rho^2 \dot \phi &= \frac{1}{\Delta}\left((\rho^2-2Mr)\frac{L_z}{\sin^2\theta}+ 2MarE \right),\nonumber\\
    (\rho^2\dot r)^2&= E^2r^4+ (a^2E^2-L_z^2-\mathscr{C})r^2+ \left[(aE-L_z)^2+\mathscr{C}\right]\,2Mr-a^2\mathscr{C}, \nonumber\\
    (\rho^2 \dot\theta)^2&=\mathscr{C}- \left(\frac{L_z^2}{\sin^2\theta}- a^2 E^2\right)\cos^2\theta,
\end{align}
where the dot identifies the derivative with respect to an affine parameter and we defined 
\begin{equation}
    E= -\xi^{\mu}_{(t)}u_{\mu}, \quad L_z = \xi^{\mu}_{(\phi)}u_{\mu}, \quad\Sigma^2= (r^2+a^2)^2- a^2\Delta \sin^2\theta,
\end{equation}
with $\xi^{\mu}_{(t,\phi)}$ being the Killing vectors associated with time translations and rotations around the $z$-axis respectively, $u_{\mu}$ is the tangent vector along the geodesic and $\mathscr{C}$ is the Carter constant~\cite{PhysRev.174.1559}.
We will work only with equatiorial orbits, both for simplicity and because the associated QNMs are the least damped, so they can be more easily measured experimentally.

Therefore we set $\theta=\pi/2$ and $\mathscr{C}=0$\footnote{We set the Carter constant to zero in the equations because it controls the latitudinal motion of the geodesic. This choice correctly describes only equatorial orbits.}, and define the \textit{impact parameter} as
\begin{equation}
    b=\frac{L_z}{E}, 
\end{equation}
to get an equation for the radial motion only
\begin{equation}
    \dot r = \frac{L_z^2}{b^2}\left(1 +\frac{a^2-b^2}{r^2}+ 2M \frac{(b-a)^2}{r^3} \right).
\end{equation}
Closed circular orbits can be obtained by setting $\dot r = \ddot r =0$
\begin{equation}
    r_{p,r}= 2M \left[1+ \cos\left(\frac 23 \arccos\left(\pm\frac{a}{M}\right)\right)\right],
\end{equation}
where $r_{p,r}$ are respectively the radii of the prograde and retrograde orbits.
The associated impact parameter is 
\begin{equation}
    b_{p,r}= -a\pm3\sqrt{Mr_{p,r}},
\end{equation}
and we can identify an orbital frequency as
\begin{equation}
    \omega_{{\rm orb, }\,p,r}=\frac{1}{b_{p,r}}.
\end{equation}
In what follows we will use the subscript $r_0, \, b_0, \, \omega_{\rm orb}$ to identify the quantity associated either with the prograde orbit $r_p$ and the retrograde orbit $r_r$, because all the calculations hold for both cases.

\section{The Penrose limit}
\label{sec:penrose_limit}
\noindent
Given a metric $g_{\mu \nu}$ and a null geodesic $\gamma$, the Penrose limit \cite{Penrose:1965rx} allows to write a plane-parallel-wave type metric (pp-wave metric) along the geodesic, describing the spacetime in a neighborhood of $\gamma$. 
By choosing a coordinate system adapted to the geodesic $\gamma$, with the 4-velocity $u_{\mu}$ along $\gamma$ and a  parallel propagated null frame $n_{\mu},\, e^{(1)}_{\mu}, \, e^{(2)}_{\mu}$ along $u_{\mu}$ with
\begin{equation}
    u_{\mu}u^{\mu}= n_{\mu}n^{\mu}= n_{\mu}e^{(i)}{}^{\mu}=u_{\mu}e^{(i)}{}^{\mu}=0,\quad n^{\mu}u_{\mu}=1, \quad e^{(i)}{}^{\mu}e^{(j)}_{\mu}=\delta^{ij},
\end{equation}
the metric can be written in the general form
\begin{equation}
\label{eq:general_ppwave}
    \dd s^2_{\rm pp} = 2 \dd u \dd v + H_{ij}(u)x^ix^j\dd u^2 + \dd x_1^2 + \dd x_2^2,
\end{equation}
in the so-called Brinkmann coordinates.

Albeit in general it is possible to construct this limit for every null geodesic, we will apply it to the closed circular orbits found in the previous sections. 
This choice greatly simplifies the calculations and at the same time it has been shown that it reproduces the QNMs spectrum \cite{Dolan_2010,Fransen:2023eqj}.

For the specific case of a Kerr black hole and focusing on equatorial circular closed orbits, we can write the null four-velocity as \cite{Igata_2019,Fransen:2023eqj}
\begin{equation}
    u_{\mu} \dd x^{\mu}=-\dd t + b_0\dd\phi,
\end{equation}
where $b_0$ is the impact parameter. 
Then the rest of the propagating null frame can be chosen as
\begin{equation}
    e^{(1)\mu}=\frac{1}{C} \left( u^{\alpha} h_{\alpha}^{\mu} + u \,u^{\mu}\right), \quad e^{(2)\mu}=\frac{1}{C} \left( u^{\alpha} f_{\alpha}^{\mu}\right)
\end{equation}
\begin{equation}
    n^{\mu}=-\frac{1}{C} e^{(1)\alpha} h_{\alpha}{}^{\mu} +\frac{1}{2C^4} \left(C_{\beta}^{\gamma}C_{\gamma\delta}u^{\beta}u^{\delta} + u^2 \,C^2 \right)u^{\mu} 
\end{equation}
with $u$ being the affine parameter along the geodesic and 
\begin{equation}
C_{\beta}^{\gamma}C_{\gamma\delta}u^{\beta}u^{\delta}= -\frac{4\Delta r_0^4}{(r_0-M)^2}, \quad C^2=\frac{4\Delta r_0^2}{(r_0-M)^2},
\end{equation}
where we also defined the Killing-Yano two-form $f$ and its dual $h$ to be
\begin{equation}
    f=-r (a\dd t - (r^2+a^2)\dd \phi)\wedge\dd \theta,\quad\quad h=r(\dd t - a \dd \phi)\wedge\dd r.
\end{equation}
It can be seen that this frame satisfies the conditions
\begin{equation}
    n_{\mu}n^{\mu}=u_{\mu}u^{\mu}=0, \quad e^{(i)\mu}e^{(j)}_{\mu}=\delta^{ij}.
\end{equation}
In particular, $e^{(1)\mu}$ points in the direction of $\dd r$ when $u=0$, being 
\begin{equation}
    e^{(1)\mu} = \left(
\begin{array}{c}
 -u \frac{\sqrt{\Delta}}{(a^2-ab +r_0^2)}\\
 \frac{r_0}{ \sqrt{\Delta} }\\
 0\\
b\,u \frac{\sqrt{\Delta}}{(a^2-ab +r_0^2)} \\
\end{array}
\right) = \left(
\begin{array}{c}
 0 \\
\frac{r_0}{ \sqrt{\Delta} }\\
 0\\
0 \\
\end{array}
\right)+ u \frac{\sqrt{\Delta}}{(a^2-ab +r_0^2)}u^{\mu},
\end{equation}
while for $e^{(2)}{}^{\mu}$ we have simply
\begin{equation}
    e^{(2)\mu} = \left(0,0,r_0,0\right).
\end{equation}

With the null frame built, the $H_{ij}(u)$ in Eq. (\ref{eq:general_ppwave}) turns out to be
\begin{equation}
    H_{ij}(u)=H_{ij}= \begin{pmatrix}
        \alpha^2&0\\
        0&-\alpha^2
    \end{pmatrix}, \qquad \alpha^2 = \frac{12 M \Delta}{r_0^3(r_0-M)^2},
\end{equation}
and therefore the metric in Eq. (\ref{eq:general_ppwave}) is explicitly written as
\begin{equation}
    \dd s^2_{\rm pp} = 2 \dd u \dd v + \alpha^2 (x_1^2-x_2^2)\dd u^2 + \dd x_1^2 + \dd x_2^2.
\end{equation}
It is immediate to check that  the Penrose limit of the Schwarzschild black hole is recovered in the $a\to 0$ limit, with \cite{Blau:2002js, Fransen:2023eqj,Kehagias:2025ntm}
\begin{equation}
    \alpha^2=\frac{1}{3M^2},
\end{equation}
and the tetrad basis becomes
\begin{equation}
    n_{\mu}\dd x^{\mu}
  \underset{a\to 0}{=}
    \frac{1}{6}(\dd t + 3\sqrt{3}M \phi),
\end{equation}
\begin{equation}
     e^{(1)}_{\mu}\dd x^{\mu} \underset{a\to 0}{=}\sqrt{3}\dd r  ,\quad e^{(2)}_{\mu}\dd x^{\mu} \underset{a\to 0}{=}3M\dd\theta.
\end{equation}
We can remark how all the calculations done in the Penrose limit are ``independent'' of the black hole spin, this information is recovered only when we match it with the asymptotic solutions. 
This is a consequence of the universality of the Penrose limit, which allows to write the metric along any geodesic as Eq. (\ref{eq:general_ppwave}).

\begin{figure}
    \centering
        \includegraphics[width=\linewidth]{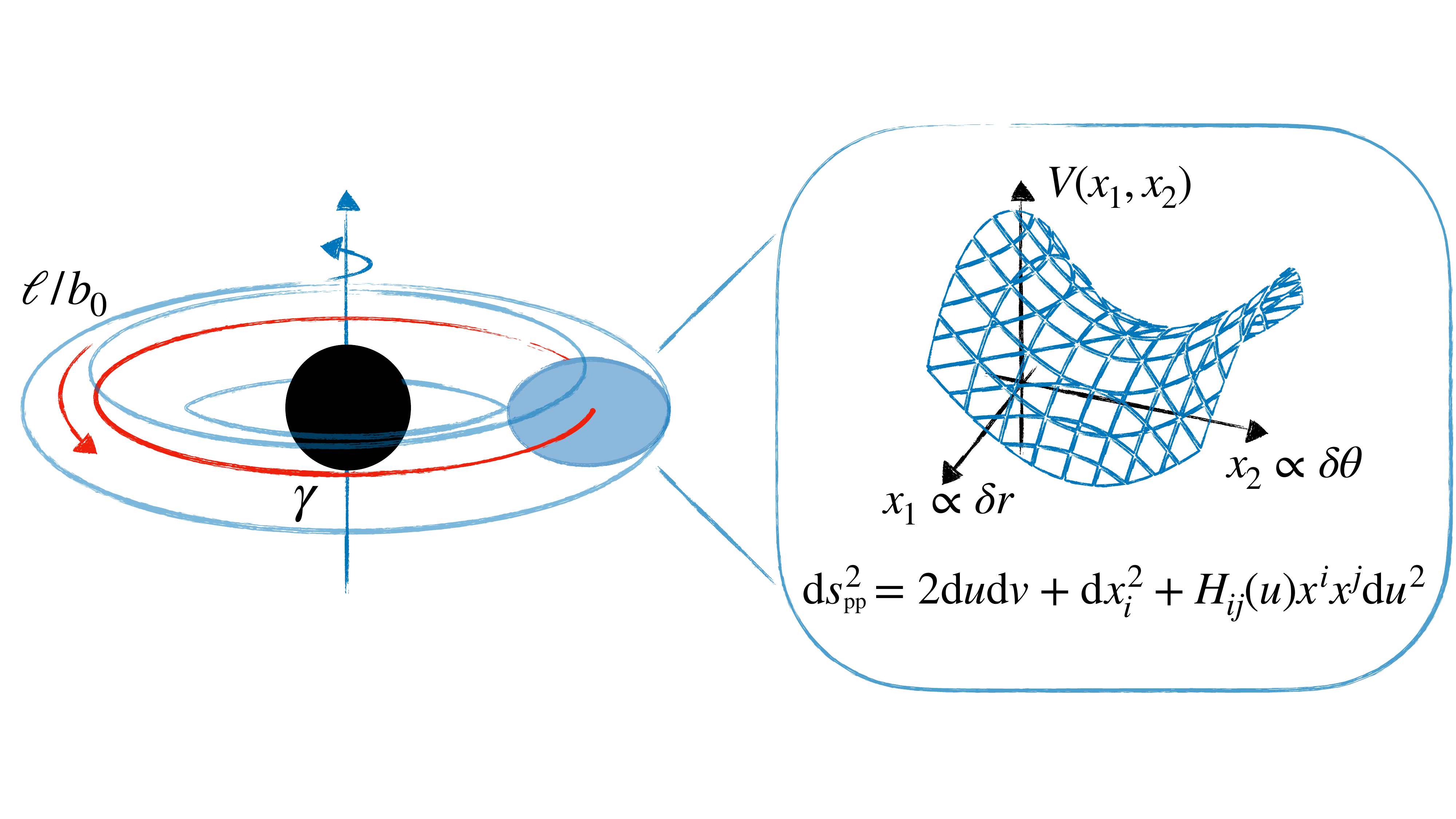}
    \caption{\it Schematic representation of the Penrose limit, highlighting the geodesic $\gamma$ and the rotation frequency of a perturbation along it $\ell/b_0$. We highlight in light blue the tubular region where the zoom is performed and what is the potential for the perturbations along the stable $(x_2)$ and unstable $(x_1)$ directions when a perturbation is added to the background $\dd s^2_{\rm {pp}}$.}
    \label{fig:schematic}
\end{figure}

\section{Linear solution within the Penrose limit}
\label{sec:linear_solution}
\noindent
\subsection{Perturbation on the pp-wave background}
\noindent
We are now ready to study how a perturbation behaves on this background.
In our previous work \cite{Kehagias:2025ntm} we focused on perturbations written in terms of Weyl scalars, however it is easier to study the metric perturbation directly and construct the Weyl scalars afterward.
This choice allows us to write the source in a more compact form and better control the contributions coming from different $\ell$'s as explained in Sec. \ref{sec:noneinear_solution}.
In  the radiation gauge (which is the same gauge where the nonlinearities are extracted numerically \cite{Loutrel:2020wbw,Ripley:2020xby}) and setting the transverse-free and traceless conditions
\begin{equation}
    \nabla_{\mu}h^{\mu\nu}{}^{(1)}=0, \quad h^{\mu}{}_{\mu}^{(1)}=0,
\end{equation}
the equations of motion for a graviton can be written as
\begin{equation}
    \nabla_{\rho}\nabla^{\rho} h_{\mu\nu}^{(2)} - 2 R^{\rho}{}_{\mu\nu\sigma}h_{\rho}^{(2)}{}^{\sigma}=0.
\end{equation}

For a pp-wave background in Brinkmann coordinates with $H_i^i=0$ we can further set the residual gauge to require $h_{v\mu}=0$, and the equations of motion become
\begin{equation}
g^{\rho\sigma}\partial_{\rho}\partial_{\sigma}h_{\mu\nu}^{(1)} + 2\delta^u_{\mu}\,\alpha^2\partial_v (h_{\nu1}^{(1)}x^1-h_{\nu2}^{(1)}x^2)+ 2\delta^u_{\nu}\,\alpha^2\partial_v (h_{\mu1}^{(1)}x^1-h_{\mu2}^{(1)}x^2)- 2\alpha^2 \delta^{u}_{\nu}\delta^{u}_{\mu} (h_{11}^{(1)}-h_{22}^{(1)})=0.
\end{equation}
From the structure of the equations we notice that a solution can be found by taking first the $(ij)$ component where the equation is the same as the scalar one and then substitute those components in the $(ui)$ equation, solve it and finally repeat the process for the $(uu)$ solution. 

The $(ij)$ components can be found by simply solving
\begin{equation}
g^{\rho\sigma}\partial_{\rho}\partial_{\sigma}h_{ij}^{(1)}=0, \quad h_{ij}^{(1)}= A_{ij}\psi(u,v,x_1,x_2).
\end{equation}
The procedure to find the solution for the scalar is described in Appendix \ref{app:linear_solution_scalar} and the solution is
\begin{equation}
    h_{ij}^{(1)}\propto  \exp\left(i p_v v+ i p_u u + i \frac{\alpha p_v}{2}(x_1^2+i x_2^2) \right),
\end{equation}
with\footnote{We choose the solution with $n_1=n_2=0$, which is the least damped solution and corresponds to the lowest ``overtone'' number. We will keep this choice through the rest of the work.}
\begin{equation}
    p_v= \ell \omega_{\rm orb}= \frac{\ell}{b_0}, \qquad p_u= \frac{\alpha}{2}(i-1),
\end{equation}
which is related to the damping of the gravitational wave signal.
By performing a similar procedure for the other components we find
\begin{equation}
    h_{\mu\nu}^{++}{}^{(1)}= A_{\mu\nu}^{++}{}^{(1)}\exp\left(i p_v v+ i p_u u + i \frac{\alpha p_v}{2}(x_1^2+i x_2^2) \right),
\end{equation}
with\footnote{This solution matches the one obtained with spin raising operators, see e.g. Refs. \cite{Adamo:2017nia, Kehagias:2025ntm}.}
\begin{equation}
    A_{\mu\nu}^{++}{}^{(1)} =\frac{i}{2}\begin{pmatrix}
        i \alpha  \left(\alpha  (x_1-x_2)^2-\dfrac{(1+i)}{p_v}\right) & 0 & -i \alpha   
(x_1-x_2) & \alpha   (x_1-x_2) \\
 0 & 0 & 0 & 0 \\
 -i \alpha   (x_1-x_2) & 0 & i  & -1 \\
 \alpha   (x_1-x_2) & 0 & -1 & -i  \\
    \end{pmatrix}.
\end{equation}

We have an equivalent solution of the form 
\begin{equation}
    h_{\mu\nu}^{--}{}^{(1)}= A_{\mu\nu}^{--}{}^{(1)}\exp\left(-i p_v v- i \bar{p}_u u - i \frac{\alpha p_v}{2}(x_1^2-i x_2^2) \right),
\end{equation}
with $A_{\mu\nu}^{--}= \bar{A}_{\mu\nu}^{++}$, which describes a wave with negative real frequency but the same damping factor. 
It was observed \cite{Ma:2024qcv, Khera:2024yrk,Bucciotti:2024jrv} that this second one is the so-called mirror mode and the two can be excited with different amplitudes, depending on the initial conditions.
We will therefore keep a different factor between the two amplitudes $A_{\mu\nu}^{--},A_{\mu\nu}^{++}$.

\subsection{Matching and asymptotic behavior at linear order}
\noindent
Alas the solution we just found holds only in a small tubular region close to the chosen prograde or retrograde orbit, as shown schematically in Fig. \ref{fig:schematic}. 
Instead the measured nonlinearity is calculated asymptotically far from the black hole.
To compare the ratio we found with the various numerical estimates, we need to match this solution to the one that propagates to $\mathcal{I}^+$.
To do this we will first construct a null perturbed tetrad for the pp-metric and for the Kerr black hole metric and we will write the perturbation in both cases in terms of the Weyl scalar $\Psi_4$,
\begin{equation}
    \Psi_4= C_{\mu\nu\rho\sigma}n^{\mu}\bar{m}^{\mu}n^{\rho}\bar{m}^{\sigma},
\end{equation}
where $C_{\mu\nu\rho\sigma}$ is the Weyl tensor. 
For the Kerr black hole we will use the standard Teukolsky equations, while for the pp-wave found from the Penrose limit we need to construct a perturbed tetrad. 
For the former we recap how to write it in the large $\ell$ limit in Appendix \ref{app:kerr_solution}, where we solve the Teukolsky equations in the eikonal limit.
At this point we will expand the result for both Weyl scalars around the photon ring, i.e. $x_1,\,x_2\to0$ for $\Psi_4$ found in the pp-wave background and $r_*\to r_0$ for the Kerr black hole background.

To find $\Psi_4$ in the pp-metric background at linear order we need to define a perturbed tetrad.
A possible choice is
\begin{align}
    l_{\mu}&=(-1,\;0,\;0,\;0);\nonumber\\
    n_{\mu}&=\left(\frac{\alpha^2}{2} (x_1-x_2^2), \;1,\;0,\;0\right )+ \epsilon\left(\frac 12 h_{uu}^{(1)},\;0,\;0,\;0\right);\nonumber\\
    m_{\mu}&=\frac{1}{\sqrt{2}}
    \left(0,\;0,\;i,\;1 \right) + \frac{\epsilon}{\sqrt{2}}
    \left(ih_{u1}^{(1)}+h_{u2}^{(1)},\;0,\; i\frac{h_{11}^{(1)}}{2}+h_{12}^{(1)} ,\;\frac 12 h_{22}^{(1)} \right);\nonumber\\
    \bar{m}_{\mu}&=\frac{1}{\sqrt{2}}
    \left(0,\;0,\;-i,\;1 \right) + \frac{\epsilon}{\sqrt{2}}
    \left(-ih_{u1}^{(1)}+h_{u2}^{(1)},\;0,\; -i\frac{h_{11}^{(1)}}{2}+h_{12}^{(1)} ,\;\frac 12 h_{22}^{(1)} \right),
\end{align}
where $\epsilon$ is a formal expansion parameter to keep track of the order.
We can now calculate the value of the Weyl scalar in the Penrose limit $\Psi_4^{(1)}|_{\rm pp}$ at first order\footnote{At large $\ell$ the contribution from $x_1,x_2$ is subleading. 
This happens because the inversion points of the classical motion where the matching should be performed scale like $x_{1,2}\sim 1/\sqrt{\ell}$, but $x_{1,2}$ always enter with a factor $\alpha$ that does not scale with $\ell$ but has the dimension of a momentum. 
Therefore by dimensional analysis we can keep only the terms independent of $x_{1,2}$. We also verified this behavior analytically.}
\begin{equation}
    \Psi_4|_{\rm pp} \simeq c_1\, \frac{p_v^2}{2}\,\exp\left(-i p_v v- i \bar{p}_u u - i \frac{\alpha p_v}{2}(x_1^2-i x_2^2) \right),
\end{equation}
where we kept only the exponential dependence on $x_1$ and $x_2$, while we sent to zero the rest like a ``stationary phase'' approximation.
From this result we can match with the approximate $\Psi_4$ from Appendix \ref{app:kerr_solution}, resulting in
\begin{equation}
    \Psi_4|_{\rm eik} \simeq c_1\frac{p_v^2}{2} e^{i S_r^{(\ell)}(r)} e^{i S_\theta^{(\ell)}(\theta)} e^{i \ell \phi}e^{-i\ell t/b_0},
\end{equation}
with $S_r^{(\ell)},\,S_\theta^{(\ell)}$ being the radial and angular part of the ``principal function'' of the Hamilton-Jacobi equation in Kerr spacetime.
After having solved the linear equations, we can turn our attention to the nonlinear solution.

\section{nonlinear solution within Penrose limit}
\label{sec:noneinear_solution}

\subsection{Perturbation with a source for $\ell_1\times \ell_2$}
\noindent
To write the nonlinear solution we first need to explore how the source is written for generic $\ell_1\times \ell_2$. 
At this point we have to remember that the source in general is written as
\begin{equation}
    S\sim h^{(1)} h^{(1)},
\end{equation}
and by expanding 
\begin{equation}
    h^{(1)}=\sum_\ell h^{(1)}_{\ell}
\end{equation}
we may write the source as 
\begin{equation}
    S\sim \sum_{\ell_1,\ell_2}h^{(1)}_{\ell_1} h^{(1)}_{\ell_2} 
    = \sum_{\ell_1\geq \ell_2} \sigma_{\ell_1,\ell_2}\,h^{(1)}_{\ell_1} h^{(1)}_{\ell_2}
\end{equation}
where we introduced the symmetry factor to avoid double counting
\begin{equation}
    \sigma_{\ell_1,\ell_2} = \begin{cases} 
      \frac 12 & \ell_1=\ell_2, \\
      1&{\rm otherwise}.
   \end{cases}
\end{equation}
In what follows we will assume $\ell_1\neq \ell_2$ and we will restore the factor $1/2$ at the end for the cases $\ell_1=\ell_2$.

Before studying the equations of motion we can compute the source and study its properties, to simplify the task of finding the solutions to the nonlinear equations. 
The full expressions for the various components of the source are written in Appendix \ref{app:source}.
However, we can highlight here some general behaviors.
Being the source quadratic in $h^{(1)}$, we can identify four possible ``phases'' 
\begin{align}
    f_{+}(u,v,x_1,x_2)&=i\left( (p_{u1}+p_{u2} )u + (p_{v1}+p_{v2})v +\frac12 \alpha (p_{v1}+p_{v2})x_1^2 \right) - \frac 12 \alpha (p_{v1}+p_{v2})x_2^2\nonumber\\
    f_{-}(u,v,x_1,x_2)&=-i\left( (\bar{p}_{u1}+\bar{p}_{u2} )u + (p_{v1}+p_{v2})v +\frac12 \alpha (p_{v1}+p_{v2})x_1^2 \right) - \frac 12 \alpha (p_{v1}+p_{v2})x_2^2\nonumber\\
    g_{+}(u,v,x_1,x_2)&=i\left( (p_{u2}-\bar{p}_{u1} )u + (p_{v2}-p_{v1})v +\frac12 \alpha (p_{v2}-p_{v1})x_1^2 \right) - \frac 12 \alpha (p_{v1}+p_{v2})x_2^2\nonumber\\
    g_{-}(u,v,x_1,x_2)&=i\left( (p_{u1}-\bar{p}_{u2} )u + (p_{v1}-p_{v2})v +\frac12 \alpha (p_{v1}-p_{v2})x_1^2 \right) - \frac 12 \alpha (p_{v1}+p_{v_2})x_2^2\nonumber,
\end{align}
with \footnote{Also here we keep only the ``ground state'' for $p_u$; for the general form see Appendix \ref{app:linear_solution_scalar}.}
\begin{equation}
    p_{vj}=\ell_{j} \,\omega_{\rm orb}, \qquad p_{uj}= \frac{\alpha}{2}(i-1), \quad j\in (1,2).
\end{equation}
From the structure of the exponents $f_{\pm}$ and $g_{\pm}$ we can see that we always have the sum of angular momenta $\ell_1+\ell_2$, associated with $p_{v1}+p_{v2}$, and present only in $f_{\pm}$, and the difference $(\ell_1-\ell_2)$, associated with $p_{v1}-p_{v2}$, and present in  $g_{\pm}$.
We will study  only the largest angular momentum $(\ell_1+\ell_2)$, and therefore,  we will consider only the term associated with the exponent $f_{\pm}$.
Another way of seeing this is the following: the metric perturbation is the sum of $++$ and~$--$ polarizations, which can be written schematically as
\begin{equation}
    h_{\ell}^{(1)}\sim (e^{i\ell v}+ e^{-i\ell v}),
\end{equation}
so that the total source schematically reads as
\begin{equation}
    S\sim h^{(1)}_{\ell_1}h^{(1)}_{\ell_2}\sim (e^{i\ell_1 v}+ e^{-i\ell_1 v})(e^{i\ell_2 v}+ e^{-i\ell_2 v})\sim e^{i(\ell_1+\ell_2)v} + e^{-i(\ell_1+\ell_2)v} + e^{i(\ell_1-\ell_2)v}+e^{-i(\ell_1-\ell_2)v}. 
\end{equation}
However, if we consider only one polarisation, e.g., $++$, the result greatly simplifies giving 
\begin{equation}
    S\sim h^{++(1)}_{\ell_1}h^{++(1)}_{\ell_2}\sim  e^{i(\ell_1+\ell_2)v},
\end{equation}
 which amounts to keeping only the term proportional to $e^{f_{+}}$. Furthermore, it is easy to see that the solution with $e^{f_{-}}$ is just the complex conjugate of the $e^{f_+}$ one.
 With this hypothesis the source greatly simplifies, giving 
\begin{equation}
S^{++}_{\mu\nu}=\tilde{S}^{++}_{\mu \nu}(x_1,x_2)\,e^{\,f_+(u,v,x_1,x_2)},
\end{equation}
where $\tilde{S}^{++}_{\mu \nu}(x_1,x_2)$ does not depend on the null coordinates $u,v$.
Furthermore, the components $h_{v\mu}$ have source terms that only depend on $g_{\pm}$, and  therefore can be set to zero at nonlinear order
\begin{equation}
    S_{v\mu}\sim 0, \quad \implies \quad h_{v\mu}^{(2)}=0.
\end{equation}
In the $\ell_1=\ell_2$ limit, the $g_\pm$-proportional terms are non-oscillatory, and can thus be neglected in the large-$\ell_1,\ell_2$  limit.

Let us now focus to the solution of the (coupled) equations
\begin{equation}
    \nabla_{\rho}\nabla^{\rho} h_{\mu\nu}^{(2)} - 2 R^{\rho}{}_{\mu\nu\sigma}h_{\rho}^{(2)}{}^{\sigma}=\tilde{S}^{++}_{\mu \nu}(x_1,x_2)\,e^{\,f_+(u,v,x_1,x_2)},
\end{equation}
which, when explicitly written, they have a structure
\begin{equation}
g^{\rho\sigma}\partial_{\rho}\partial_{\sigma} h_{uu}^{(2)} + 4 \alpha^2 \partial_v (h_{u1}^{(2)} x_1- h_{u2}^{(2)}x_2) - 2\alpha^2 (h_{11}^{(2)}- h_{22}^{(2)})  = \tilde{S}^{++}_{uu}(x_1,x_2)\,e^{\,f_+(u,v,x_1,x_2)},
\end{equation}
\begin{equation}
g^{\rho\sigma}\partial_{\rho}\partial_{\sigma} h_{ui}^{(2)} + 2 \alpha^2 \partial_v (h_{1i}^{(2)} x_1- h_{2i}^{(2)}x_2) = \tilde{S}^{++}_{ui}(x_1,x_2)\,e^{\,f_+(u,v,x_1,x_2)},
\end{equation}
\begin{equation}
g^{\rho\sigma}\partial_{\rho}\partial_{\sigma} h_{ij}^{(2)} = \tilde{S}^{++}_{ij}(x_1,x_2)\,e^{\,f_+(u,v,x_1,x_2)}. 
\end{equation}
We will follow in the sequence the same steps as in the linear case, solving first for the component $(i,j)$, then for the $(u,i)$ and finally for the $(u,u)$, but now  in the presence of a source.
Keeping only the least damped state we are led to the following solution for the nonlinear equations
\begin{equation}
    h_{\mu\nu}^{++}{}^{(2)}= A_{\mu\nu}^{++}{}^{(2)} \frac{(p_{v1}+p_{v2})^2}{4p_{v1}p_{v2}}e^{\,f_+(u,v,x_1,x_2)},
\end{equation}
with
\begin{equation}
    A_{\mu\nu}^{++}{}^{(2)} =\begin{pmatrix}
        i \alpha  \left(\alpha  (x_1-x_2)^2-\dfrac{(1+i)}{(p_{v1}+p_{v2})}\right) & 0 & -i \alpha   
(x_1-x_2) & \alpha   (x_1-x_2) \\
 0 & 0 & 0 & 0 \\
 -i \alpha   (x_1-x_2) & 0 & i  & -1 \\
 \alpha   (x_1-x_2) & 0 & -1 & -i  \\
    \end{pmatrix}.
\end{equation}
It can be checked that the solution for  $h_{\mu\nu}^{--}{}^{(2)}$ is just the complex conjugate of the solution written above. 
Having derived the metric up to second order in perturbations, we proceed to compute the corresponding tetrad and Weyl scalars, facilitating comparison with the asymptotic free-field solutions.

\subsection{Matching and asymptotic behavior at nonlinear order}
\noindent
As in Section \ref{sec:linear_solution}, we need to find the Weyl scalar $\Psi_4$ in order to perform the matching.
It is important to begin this section with the following remark.
The source that enters in the second order Teukolsky equation, can be written schematically as
\begin{equation}
    S\sim f(r)\left(\Psi(r)\right)^2, 
\end{equation}
where the function $f(r)$ is peaked near the light ring, as has been noticed in \cite{Nakano:2007cj,Bucciotti:2024jrv}. 
Therefore, in line with our ``stationary phase'' approximation, we will make the assumption that the source is localized exactly at the light ring.
We assume also that the gravitational wave propagates freely after exiting the light ring, allowing us to match the solution at nonlinear level with the homogeneous solution, greatly simplifying the calculations.
This approximation led to a fairly accurate estimate in a previous work \cite{Kehagias:2025ntm}.

The tetrad up to second order can be written as
\begin{align}
    \delta l^{(2)}_{\mu}=&(0,\;0,\;0,\;0);\nonumber\\
    \delta n_{\mu}^{(2)}=&\frac{\epsilon^2}{2}\left( h_{uu}^{(2)}- \left(h^{(2)}_{u1}\right)^2- \left(h^{(2)}_{u2}\right)^2,\;0,\;0,\;0\right);\nonumber\\
   \delta m^{(2)}_{\mu}=& \frac{\epsilon^2}{2\sqrt{2}}\Bigg{(}i (-h^{(1)}_{11} h^{(1)}_{u1}-2 h^{(1)}_{12} h^{(1)}_{u2}+i h^{(1)}_{22} h^{(1)}_{u2}+2 h^{(2)}_{u1}-2 i h^{(2)}_{u2}),0,\nonumber\\
   & -\frac{i}{4} \left(\left(h^{(1)}_{11}\right)^2+4 \left(h^{(1)}_{12}\right)^2-4 i h^{(1)}_{12} h^{(1)}_{22}-4 h^{(2)}_{11}+8 i h^{(2)}_{12}\right),\frac{4 h^{(2)}_{22}-\left(h^{(1)}_{22}\right)^2}{4}\Bigg{)};\nonumber\\
    \delta \bar{m}^{(2)}_{\mu}=& \frac{\epsilon^2}{2\sqrt{2}}  \Bigg{(}-i (-h^{(1)}_{11} h^{(1)}_{u1}-2 h^{(1)}_{12} h^{(1)}_{u2}-i h^{(1)}_{22} h^{(1)}_{u2}+2 h^{(2)}_{u1}+2 i h^{(2)}_{u2}),\;0,\nonumber\\
    &\frac{i}{4} \left(\left(h^{(1)}_{11}\right)^2+4 \left(h^{(1)}_{12}\right)^2+4 i h^{(1)}_{12} h^{(1)}_{22}-4 h^{(2)}_{11}-8 i h^{(2)}_{12}\right),\;\frac{4 h^{(2)}_{22}-\left(h^{(1)}_{22}\right)^2}{4}\Bigg{)}.
\end{align}
From the above  equations and the expansion of $C_{\mu\nu\rho\sigma}^{(2)}$ at second order, we can evaluate the Weyl scalar $\Psi_4^{(2)}$, which when expanded around $x_1,\,x_2\sim 0$, turns out to be
\begin{equation}
\label{eq:psi_4_nonlin_amp}
    \Psi_4^{(2)}\simeq c_1^2 \frac{i \sigma_{\ell_1,\ell_2}  (p_{v1}+p_{v2})^4-\left(i \alpha ^4+p_{v1}^3 p_{v2}+p_{v1} \
p_{v2}^3\right)}{4 p_{v1} p_{v2}} e^{f_-(u,v,x_1,x_2)} \;+ \bar{c}_1^2\frac{i \alpha ^4}{4 p_{v1} p_{v2}} e^{f_+(u,v,x_1,x_2)}.
\end{equation}
We  have kept in Eq. (\ref{eq:psi_4_nonlin_amp}) only the terms proportional to $f_{\pm}$, whereas $\sigma_{\ell_1,\ell_2}$ takes into account the symmetry factor. 
Notice that the mirror mode associated with $f_{+}$ is suppressed in the limit of large $\ell$, consistently with  Ref. \cite{Ma_2024}.
Therefore, form now on we will keep only the term proportional to $f_{-}$.

Following the same matching procedure as the linear case, we can write at leading order in $\ell$
\begin{equation}
    \Psi_4^{(2)}|_{\rm eik} \simeq c_2 e^{i S_r^{\ell_{\rm tot}}(r)} e^{i S_\theta^{\ell_{\rm tot}}(\theta)} e^{i \ell_{\rm tot} \phi}e^{-i\ell_{\rm tot}\ell t/b_0},\quad \ell_{\rm tot}=\ell_1+\ell_2
\end{equation}
with
\begin{equation}
    c_2=c_1^2 \frac{i \sigma_{\ell_1,\ell_2}  (p_{v1}+p_{v2})^4-\left(i \alpha ^4+p_{v1}^3 p_{v2}+p_{v1} \
p_{v2}^3\right)}{4 p_{v1} p_{v2}},
\end{equation}
where the approximation
\begin{equation}
\label{eq:approximate_eigenfreq}
    \omega_{\ell_1, \ell_1, 0}+ \omega_{\ell_2, \ell_2, 0} \simeq \frac{\ell_1 + \ell_2}{b_0} -i \alpha \simeq \omega_{\ell_1+\ell_2, \ell_1+\ell_2,0},
\end{equation}
has been employed, consistent with leading-order behavior in $\ell$. 
Alas going to the next-to-leading order approximation would be very complex because asymptotically we would need terms $\mathcal{O}(\ell^0)$ and the equations of motion for the free wave would not have eigenfrequencies respecting the approximation of Eq. (\ref{eq:approximate_eigenfreq}).
We have now all the ingredients to write the nonlinear amplitude ratio and the phase.

\section{Amplitude and phase ratios}
\label{sec:amp_phase}
\noindent
Having written the linear and nonlinear solution for $\Psi_4$ at $\mathcal{I}^+$ we can now calculate the ratio between the nonlinear solution and the linear solution squared, extracting the amplitude and the phase difference with the additional symmetry assumption, i.e. that the amplitude for $A^{++}$ and $A^{--}$ is the same. 
If this is the case, the dependence on the amplitude cancels, resulting in a complex number.
However in the general case where the symmetry condition is not respected the nonlinear ratio  dependent on initial conditions, as argued by Refs. \cite{Ma:2024qcv, Khera:2024yrk,Bucciotti:2024jrv}.
Our formulas can be generalized for different amplitudes. 
This requires keeping track of the functions in front of all the possible exponentials $\exp(g_{\pm}),\,\exp(f_{\pm})$, and thus complicating the calculations. 

The amplitude ratio is defined as
\begin{equation}
\mathcal{R}_{\ell_1\times \ell_2}= \frac{\mathcal{A}^{\ell_1\times \ell_2}_{\ell_1+\ell_2\,\ell_1+\ell_2}}{\mathcal{A}_{\ell_1\,\ell_1}\;\mathcal{A}_{\ell_2\,\ell_2}},
\end{equation}
where the amplitude $\mathcal{A}_{\ell_i\,\ell_i}$ is the one that appears in the TT gauge expansion of the gravitational wave\footnote{We are considering the sum only on the fundamental overtone $n=0$, which is least damped QNM.}
\begin{equation}
    h_+ -ih_{\times} = \frac{M}{r}\sum_{\ell m}\mathcal{A}_{\ell m} \,{}_{-2}S_{\ell m}(\theta) e^{-i\omega_{\ell m} (t-r)}e^{i m\phi}.
\end{equation}
To connect the amplitude $\mathcal{A}_{\ell m}$ with the amplitude of $\Psi_4$ we can use the relation valid asymptotically
\begin{equation}
    \mathcal{A}_{\ell m} = -\frac{4}{\omega^2_{\ell m}}\mathcal{A}^{\Psi}_{\ell m}.
\end{equation}
The normalization of the asymptotic tetrad is chosen to have $\Psi_4=\Psi_0^*$.
Finally, to write the amplitude correctly we need to project the phase $\exp(i S_{\theta}(\theta))$ onto the correspondent spin-weighted spheroidal harmonic\footnote{In particular in the limit of large $\ell$ it is possible to drop the spin dependence of ${}_{-2}S_{\ell m}$ and write it just as a scalar spheroidal harmonic.}. 
The nonlinear ratio is therefore

\begin{equation}
    \mathcal{R}_{\ell_1\times \ell_2}=\left( \frac{\ell_1^3 \ell_2+\ell_1 \ell_2^3-i \sigma_{\ell_1,\ell_2}(\ell_1+\ell_2)^4+i(\alpha^4/\omega^4)}{8 \ell_1 \ell_2 (\ell_1+\ell_2)^2}\right)\frac{\mathcal{C}_{\ell_1+\ell_2}}{\mathcal{C}_{\ell_1}\mathcal{C}_{\ell_2}},
\end{equation}
with
\begin{equation}
    \mathcal{C}_{\ell} = 2\pi\int_0^{\pi} \,{}_{-2}S_{\ell\ell}(\theta)\,e^{i S^{(\ell)}_{\theta}(\theta)}\sin \theta \,\dd \theta.
\end{equation}
The dependence in $m$ of ${}_{-2}S_{\ell m}(\theta)$ is fixed by the integral along $\phi$, and the projection is maximized when the angular momentum value of the spheroidal harmonic $\ell$ matches the one appearing in $S^{(\ell)}_{\theta}$.

In the ratio we kept the subleading term in $\alpha$, because it encodes the dependence w.r.t the black hole spin.
This is not consistent with our choice of retaining only the leading order in $\ell$,  However this dependence enters at the level of the nonlinearity located at the light ring, while the subleading terms we ignore are all related to the precision of the matching of the free wave.
Therefore, the dependence on $\alpha$, which encodes the spin of the black hole, reflects the light ring physics. 
The behavior at large black hole spins will get larger corrections from the next-to-leading order eikonal approximation.\\

\begin{figure}
     \centering
     \subfloat[]{
          \includegraphics[width=0.44\linewidth]{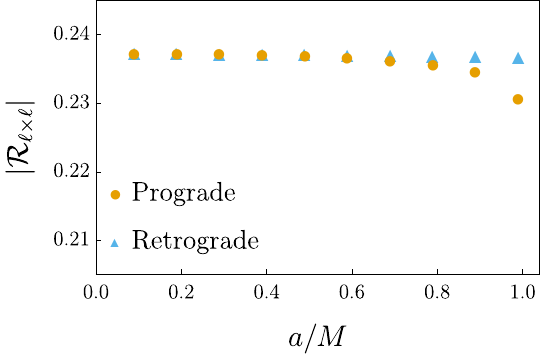}} \subfloat[]{\includegraphics[width=0.46\linewidth]{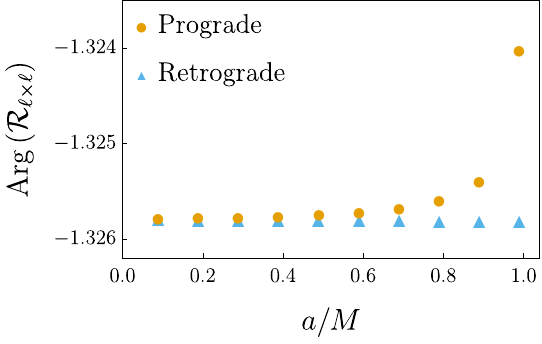}}
      \caption{\it Amplitude and phase of the nonlinear ratio $\mathcal{R}_{\ell\times \ell}$ as a function of the adimensional black hole spin $a/M$. We show both the retrograde and prograde orbits and for $\ell_1=\ell_2=\ell=10$.}
       \label{fig:Rll_v_a}
\end{figure}

In Fig. \ref{fig:Rll_v_a} we show the results for amplitude and phase of the nonlinear ratio for $\ell_1=\ell_2=10$. 
We notice an analogous behavior with higher values of $\ell$, where the amplitude slightly decreases as we approach extremality $a/M\to1$, similarly to what found by Refs. \cite{Ma_2024,Khera:2024yrk,Zhu:2024rej,Redondo-Yuste:2023seq}.
The phase instead seem to stabilize around $\sim\exp(-2i\pi/5)$ for higher values of $\ell$. 
The retrograde orbit is in general less sensitive to the black hole spin, because when the latter increases the impact parameter $b_p$ of the prograde orbit decreases sharply (and its frequency consequently increases) while for the retrograde orbit $b_r$ increases slowly, thus resulting in more stable values for the latter case.

\begin{figure}
     \centering
\subfloat[]{          \includegraphics[width=0.44\linewidth]{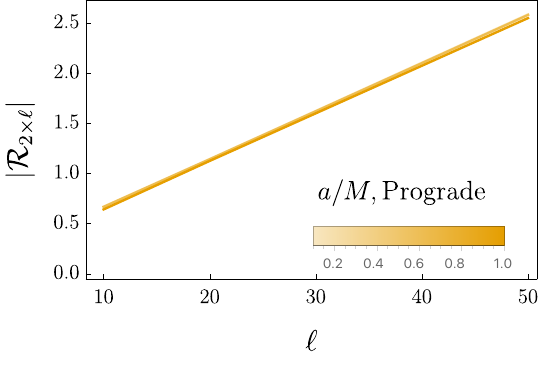}} \subfloat[]{\includegraphics[width=0.45\linewidth]{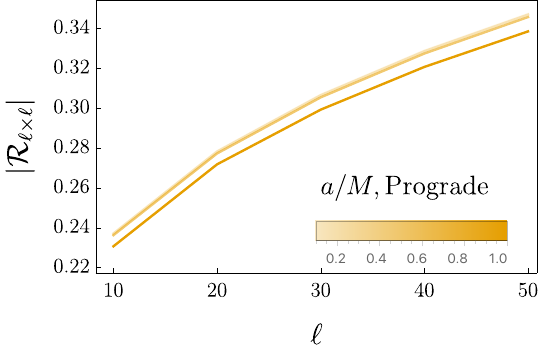}}
      \caption{\it Dependence with respect to $\ell$ of the nonlinear ratio for the two cases $\mathcal{R}_{2\times\ell} $ and $\mathcal{R}_{\ell\times\ell}$.}
      \label{fig:R_v_l}
\end{figure}

In Fig. \ref{fig:R_v_l} we plotted how the amplitude of the ratios $\mathcal{R}_{2\times \ell}$ and $\mathcal{R}_{\ell\times \ell}$ depend on $\ell$. 
For the former we have a linear scaling with little dependence on the black hole spin, which increases as it approaches its extremal values.
The slope in this case is $\sim 0.08\ell$, to be confronted to the second-order numerical result $\sim 0.11\ell$ of  Ref. \cite{Bucciotti:2025rxa} and therefore in rather good agreement despite the fact that $\ell=2$ is outside of the eikonal regime.
For the amplitude of $\mathcal{R}_{\ell\times \ell}$ we notice a residual dependence on $\ell$, scaling as $\ell^{1/4}$. 
This is consistent with the results found for $\mathcal{R}_{2\times 2}$ in the Schwarzschild case, where the result depends on the Gaunt integral of three spherical harmonics (see e.g. Refs. \cite{Brizuela_2009, Bucciotti:2024jrv, Bucciotti:2024zyp}), which grows as $\ell^{1/4}$.  
We highlight that the spin dependence is very mild, especially in the $2\times\ell $ case.

\section{Conclusions}
\label{sec:conclusions}
\noindent
In this work we studied the spin dependence of the eikonal regime of QNMs, including  nonlinearities.
In particular we focused on the nonlinear regime of a QNM in the Penrose limit, zooming close to a circular closed orbit and solving the second order Einstein equations. 
We then proceed to match this solution to an eikonal vacuum solution for both the linear and the nonlinear case, finding an estimate for the nonlinearity amplitude and phase in the large $\ell$ regime.

With this work we build upon current studies of nonlinearity, providing analytic results for a specific simple case, which can be tested with current numerical studies.
Our work can be extended in multiple directions.
First it is possible to include next-to-leading order corrections to all the results, obtaining more precise results within the working assumption of a localized source.
Another possible extension is to keep track of the phases $g_{\pm}$ to better understand the numerical and semi-analytical results in the literature with this simple formalism.

Furthermore a study of the complete nonlinear equations in the large $\ell$ limit without the aid of the Penrose limit can be a useful check of the result, exploring its limits and supplement additional testing for the assumptions.
We plan to expand on these points in future works.

\section*{Acknowledgements}
\noindent
We thank C. Beadle, F. Bernardo, A. Ianniccari and M. Pijnenburg for useful discussions.
A.R. acknowledges support from the Swiss National Science Foundation (project number CRSII5 213497) and from the Boninchi Foundation for the project “PBHs in the Era of GW Astronomy”. 
The work of D.P. is supported by the Swiss National Science Foundation under grants no. 200021-205016 and PP00P2-206149.

\appendix
\setcounter{equation}{0}
\setcounter{section}{0}
\setcounter{table}{0}
\makeatletter
\renewcommand{\theequation}{A\arabic{equation}}

\section{Source}
\label{app:source}
\noindent
We report below all the components of the  source term  in the Penrose limit:
\begin{align}
S_{uu}=&\frac{\alpha^2(p_{v1}+p_{v2})^2}{4p_{v1}p_{v2}} \left[e^{f_+(u,v,x_1,x_2)}\left(\alpha (1+i)(p_{v1}+p_{v2})(x_1-x_2) -2i\right) + \right.\nonumber\\
    &\left.e^{f_-(u,v,x_1,x_2)}\left( \alpha(1-i)(p_{v1}+p_{v2})(x_1-x_2) +2i\right)\right]+\nonumber\\
    &e^{g_-(u,v,x_1,x_2)}\left\{-p_{u2}^2 -\bar{p}_{u1}^2+p_{u2} (\bar{p}_{u1}+\alpha  (\alpha  (x_1-x_2) (p_{v1} x_1-x_2 (p_{v1}+2 p_{v2}))-(1-i)))+\right.\nonumber\\
    &\alpha  \bar{p}_{u1} (\alpha  (x_1-x_2) (p_{v2} x_1-x_2 (2 p_{v1}+p_{v2}))-(1+i))+\nonumber\\
    &\frac{1}{4 p_{v1} p_{v2}}\left[-2 \alpha ^2 \left(p_{v1}^2+p_{v2}^2\right)-4 \alpha ^4 p_{v1} p_{v2} x_2^2 \left(p_{v1}^2+p_{v2}^2\right) (x_1-x_2)^2+\right.\nonumber\\
    &(1+i) \alpha ^3 (x_1-x_2) \left(x_1 (p_{v1}+i p_{v2}) \left(p_{v1}^2-4 i p_{v1} p_{v2}-p_{v2}^2\right)-\right.\nonumber\\
    &\left.\left.\left.x_2 (p_{v1}+p_{v2}) \left(p_{v1}^2+(3-3 i) p_{v1} p_{v2}-i p_{v2}^2\right)\right)\right]\right\}+\nonumber\\
&e^{g_+(u,v,x_1,x_2)}\left\{-p_{u1}^2-\bar{p}_{u2}^2+p_{u1} (\bar{p}_{u2}+\alpha  (\alpha  (x_1-x_2) (p_{v2} x_1-x_2 (2 p_{v1}+p_{v2}))-(1-i)))+\right.\nonumber\\
&\alpha  \bar{p}_{u2} (\alpha  (x_1-x_2) (p_{v1} x_1-x_2 (p_{v1}+2 p_{v2}))-(1+i))+\nonumber\\
&\frac{1}{4 p_{v1} p_{v2}}\left[-2 \alpha ^2 \left(p_{v1}^2+p_{v2}^2\right)-4 \alpha ^4 p_{v1} p_{v2} x_2^2 \left(p_{v1}^2+p_{v2}^2\right) (x_1-x_2)^2+\right.\nonumber\\
&(1+i) \alpha ^3 (x_1-x_2) \left(i x_2 (p_{v1}+p_{v2}) \left(p_{v1}^2+(3+3 i) p_{v1} p_{v2}+i p_{v2}^2\right)+\right.\nonumber\\
&\left.\left.\left.x_1 \left(-i p_{v1}^3+3 p_{v1}^2 p_{v2}-3 i p_{v1} p_{v2}^2+p_{v2}^3\right)\right)\right]\right\},\nonumber\\
&
\end{align}

\begin{align}
    S_{u1}=&-\frac{\alpha^2(p_{v1}+p_{v2})^3(x_1-x_2)}{4p_{v1}p_{v2}}\left((1+i)e^{f_+(u,v,x_1,x_2)}+(1-i)e^{f_-(u,v,x_1,x_2)}\right)+\nonumber\\
    &e^{g_-(u,v,x_1,x_2)}\left\{-\frac{1}{2} p_{u2} x_1 (p_{v1}+2 p_{v2})+p_{u2} x_2 (p_{v1}+p_{v2})-\frac{1}{2} \bar{p}_{u1} x_1 (2 p_{v1}+p_{v2})+\nonumber\right.\\
    &\bar{p}_{u1} x_2 (p_{v1}+p_{v2})+\frac{\left(1+i\right) \alpha  (x_1-x_2)}{4p_{v1} p_{v2}} \left[-p_{v1}^3-(2-3 i) p_{v1}^2 p_{v2}-\right.\nonumber\\
    &\left.\left.(2-2 i) \alpha  p_{v1} p_{v2} x_2 \left(p_{v1}^2+p_{v2}^2\right) (x_1-x_2)-(3-2 i) p_{v1} p_{v2}^2+i p_{v2}^3\right]\right\}+\nonumber\\
&e^{g_+(u,v,x_1,x_2)}\left\{-\frac{1}{2} \alpha  x_1 (p_{u1} (2 p_{v1}+p_{v2})+\bar{p}_{u2} (p_{v1}+2 p_{v2}))+\right. \nonumber\\
&\alpha  x_2 (p_{u1}+\bar{p}_{u2}) (p_{v1}+p_{v2})- \left(\alpha ^3 x_2 \left(p_{v1}^2+p_{v2}^2\right) (x_1-x_2)^2\right)-\nonumber\\
&\left.\frac{\left(1-i\right) \alpha ^2 }{4p_{v1} p_{v2}}\left(p_{v1}^3+(2+3 i) p_{v1}^2 p_{v2}+(3+2 i) p_{v1} p_{v2}^2+i p_{v2}^3\right) (x_1-x_2)\right\},\nonumber\\
&
\end{align}

\begin{align}
    S_{u2}=&\frac{ \alpha ^2 (p_{v1}+p_{v2})^3 (x_1-x_2)}{4p_{v1} p_{v2}}\left( \left(1+i\right)e^{f_-(u,v,x_1,x_2)} + (1-i)e^{f_-(u,v,x_1,x_2)}\right)+ \nonumber\\
    &e^{g_+(u,v,x_1,x_2)}\frac{\left(1+i\right) \alpha  }{4p_{v1} p_{v2}}\left\{ (1+i) p_{v1} p_{v2} x_2 (\bar{p}_{u2} p_{v1}-p_{u1} p_{v2})+\right.\nonumber\\
    &\left.\alpha  (p_{v1}+i \
p_{v2}) (p_{v1}-i p_{v2})^2 (x_2-x_1)\right\}-\nonumber\\
&e^{g_-(u,v,x_1,x_2)}\frac{\left(1+i\right) \alpha}{4p_{v1} p_{v2}} \left\{ (1+i) p_{v1} p_{v2} x_2 (p_{u2} p_{v1}-\bar{p}_{u1} p_{v2})+\right.\nonumber\\
&\left.\alpha  (p_{v2}+i \
p_{v1}) (p_{v1}+i p_{v2})^2 (x_2-x_1)\right\},\nonumber\\
&
\end{align}

\begin{align}
    S_{11}= &\frac{\alpha \,(p_{v1}+p_{v2})^3 }{4p_{v2} p_{v1}} \left((1+i)e^{f_{+}(u,v,x_1,x_2)} + (1-i)e^{f_{-}(u,v,x_1,x_2)}+ \right.\nonumber\\
    &\frac{e^{\,g_{+}(u,v,x_1,x_2)} }{\alpha \,(p_{v1}+p_{v2})^3 } \left\{4 p_{v1} p_{v2} (p_{u2} p_{v1}+\bar{p}_{u1} p_{v2})+\alpha (1+i) (p_{v1}+i p_{v2}) \left(p_{v1}^2-4 i p_{v1} p_{v2}-p_{v2}^2\right)-\right.\nonumber\\
    &\left.4 \alpha ^2 p_{v1} p_{v2} \left(p_{v1}^2 (x_1-x_2)^2-p_{v1} p_{v2} x_1^2+p_{v2}^2 (x_1-x_2)^2\right)\right\}+\nonumber
    \\
   & \frac{e^{\,g_{-}(u,v,x_1,x_2)} }{\alpha \,(p_{v1}+p_{v2})^3 } \left\{4 p_{v1} p_{v2} (p_{u1} p_{v2}+\bar{p}_{u2} p_{v1})+\alpha(1-i) (p_{v1}-ip_{v2}) \left(p_{v1}^2 + 4i p_{v1}p_{v2}- p_{v2}^2\right)-\right. \nonumber\\
   &\left.\left.4 \alpha ^2 p_{v1} p_{v2} \left(p_{v1}^2 (x_1-x_2)^2-p_{v1} p_{v2} x_1^2+p_{v2}^2 (x_1-x_2)^2\right)\right\}\right),\nonumber\\
   &
\end{align}

\begin{align}
    S_{22}= &\frac{\alpha \,(p_{v1}+p_{v2})^3 }{4p_{v2} p_{v1}} \left(-(1+i)e^{f_{+}(u,v,x_1,x_2)} - (1-i)e^{f_{-}(u,v,x_1,x_2)}+ \right.\nonumber\\
    &\frac{e^{\,g_{+}(u,v,x_1,x_2)} }{\alpha \,(p_{v1}+p_{v2})^3 } \left\{4 p_{v1} p_{v2} (p_{u2} p_{v1}+\bar{p}_{u1} p_{v2})-\alpha (1+i) (p_{v1}+i p_{v2}) \left(p_{v1}^2-p_{v2}^2\right)+4 \alpha ^2 p_{v1}^2 p_{v2}^2x_2^2 \right\}+\nonumber
    \\
   &\left. \frac{e^{\,g_{-}(u,v,x_1,x_2)} }{\alpha \,(p_{v1}+p_{v2})^3 } \left\{4 p_{v1} p_{v2} (p_{u1} p_{v2}+\bar{p}_{u2} p_{v1})-\alpha(1-i) (p_{v1}-ip_{v2}) \left(p_{v1}^2 - p_{v2}^2\right)+4 \alpha ^2 p_{v1}^2 p_{v2}^2x_2^2\right\}\right), \nonumber\\
   &
\end{align}

\begin{align}
    S_{12}= &\frac{\alpha (p_{v1}+p_{v2})^3}{4p_{v2} p_{v1}} \left(-(1-i)e^{f_{+}(u,v,x_1,x_2)} - (1+i) e^{f_{-}(u,v,x_1,x_2)}+ \right.\nonumber\\
    &\frac{e^{\,g_{+}(u,v,x_1,x_2)} }{\alpha \,(p_{v1}+p_{v2})^3 } \left[\alpha (1+i)(ip_{v1}+p_{v2})(p_{v2}-ip_{v1})^2 \right]+\nonumber
    \\
   &\left. \frac{e^{\,g_{-}(u,v,x_1,x_2)} }{\alpha \,(p_{v1}+p_{v2})^3 } \left[\alpha (1+i)(p_{v1}-ip_{v2})(p_{v1}-ip_{v2})^2\right]\right).\nonumber\\
   &
\end{align}
we remind that
\begin{align}
    f_{+}(u,v,x_1,x_2)&=i\left( (p_{u1}+p_{u2} )u + (p_{v1}+p_{v2})v +\frac12 \alpha (p_{v1}+p_{v2})x_1^2 \right) - \frac 12 \alpha (p_{v1}+p_{v2})x_2^2,\nonumber\\
    f_{-}(u,v,x_1,x_2)&=-i\left( (\bar{p}_{u1}+\bar{p}_{u2} )u + (p_{v1}+p_{v2})v +\frac12 \alpha (p_{v1}+p_{v2})x_1^2 \right) - \frac 12 \alpha (p_{v1}+p_{v2})x_2^2,\nonumber\\
    g_{+}(u,v,x_1,x_2)&=i\left( (p_{u2}-\bar{p}_{u1} )u + (p_{v2}-p_{v1})v +\frac12 \alpha (p_{v2}-p_{v1})x_1^2 \right) - \frac 12 \alpha (p_{v1}+p_{v2})x_2^2,\nonumber\\
    g_{-}(u,v,x_1,x_2)&=i\left( (p_{u1}-\bar{p}_{u2} )u + (p_{v1}-p_{v2})v +\frac12 \alpha (p_{v1}-p_{v2})x_1^2 \right) - \frac 12 \alpha (p_{v1}+p_{v_2})x_2^2\nonumber,\\
    &
\end{align}
and 
\begin{equation}
    p_v=\frac{\ell}{b_0}, \quad p_{u}= i\alpha \left(n_1+\frac 12\right)- \alpha \left(n_2+\frac 12\right).
\end{equation}

\section{Linear solution for a scalar field in a pp-wave background}
\label{app:linear_solution_scalar}
\renewcommand{\theequation}{B\arabic{equation}}
\noindent
In this appendix we recap the procedure to find the solution for a scalar field in the pp-wave background.
We will follow mostly Ref. \cite{Fransen:2023eqj}.
For a scalar field $\psi$, the equation of motion is written as 
\begin{equation}
     (2\partial_u\partial_v - \alpha^2(x_1^2-x_2^2)\partial_v^2 + \partial_1^2+\partial_2^2)\psi=0.
\end{equation}
and it can be solved by  the  ansatz 
\begin{equation}
    \psi=  \phi_1(x_1)\phi_2(x_2) \,e^{i p_v v + i p_u u}.
\end{equation}
This for of the solution leads to a differential equation for $\phi_1(x_1)\phi_2(x_2)$, which can be further separated, obtaining 
\begin{align}
    \phi_1''(x_1) +\alpha^2 p_v^2\,x_1^2 \phi_1(x_1) = (p_up_v + \xi ) \phi_1(x_1),\\
    \phi_2''(x_2) - \alpha^2 p_v^2\,x_2^2 \phi_2(x_2) = (p_up_v - \xi) \phi_2(x_2),
\end{align}
with $\xi$ being the separation constant.
Performing the following change of variables
\begin{equation}
\label{eq:variable_change_to_ho}
    y_1 = e^{-i\pi/4}\sqrt{2\alpha p_v}\,  x_1, \qquad y_2=\sqrt{2\alpha p_v}\, x_2,
\end{equation}
and defining
\begin{equation}
\label{eq:splitting_constants}
    i\frac{p_u p_v +\xi}{2\alpha p_v} = - \left( n_1 + \frac 12\right), \qquad \frac{p_u p_v - \xi}{2\alpha p_v} = - \left( n_2 + \frac 12\right),
\end{equation}
we bring both equations in the form
\begin{equation}
\label{eq:harmonic_oscillator}
    \phi_i'' -\frac 14 y_i^2 \phi_i = -\left( \frac 12 +n_i\right)\phi_i, \qquad i=1,2.
\end{equation}
To find QNMs we need to set
\begin{align}
    &e^{i\alpha p_v x_1^2/2},~\,\qquad  x_1\to \pm \infty, \\
   & e^{-\alpha p_v x_2^2/2},\qquad  x_2\to \pm \infty,
\end{align}
which are the appropriate boundary conditions for QNMs, being purely outgoing and purely ingoing in $x_1\sim \delta r$ direction and stable around $x_2\sim \delta \theta$.
To ensure this behavior it is necessary to choose the decaying solution for both cases in the variables $y_i$, because
\begin{equation}
    e^{-y_1^2/4}\to  e^{i \alpha p_v x_1^2/2}, \qquad e^{-y_2^2/4}\to e^{- \alpha p_v x_2^2/2}.
\end{equation}
This condition also fixes the spectrum and the splitting constant $\xi$. Considering $n_i\in \mathbb{N}$ we get from Eq. (\ref{eq:splitting_constants}) 
\begin{equation}
    p_u= \alpha \left[ i \left( n_1+\frac 12\right)- \left( n_2+\frac 12\right)\right], 
    \end{equation}
    and 
    \begin{equation}
    \xi= \alpha p_v\left[ i \left( n_1+\frac 12\right)+ \left( n_2+\frac 12\right)\right].
\end{equation}
Therefore we can write the frequencies of the QNMs in the Penrose limit for a Kerr black hole as \cite{Fransen:2023eqj}
\begin{equation}
    \omega_{\ell\, \ell\, n} = \frac{\ell}{b_0}+ i\omega_{\rm prec}\left(n_1+\frac 12\right)+\omega_{\rm prec}\left(n_2+\frac 12\right),
\end{equation}
\begin{equation}
    \omega_{\rm prec}^2= \alpha^2 \frac{(r_0-M)^2}{(r_0+3M)^2}.
\end{equation}
However in the work we kept only the least damped mode  $n_1=n_2=0$. With this choice the real part will be dominated by $\ell$ while the decaying one will be fixed by $i\omega_{\rm prec}\propto i \alpha$.

\section{Eikonal limit of equations of motion for \texorpdfstring{$\Psi_4$}{Psi4} in Kerr black hole background}
\label{app:kerr_solution}
\renewcommand{\theequation}{C\arabic{equation}}
\noindent
Here  we recap how to find the solution for the Weyl scalar $\Psi_4$ in the eikonal limit.
Following  Refs. \cite{Hadar:2022xag,Dolan_2010}, we start from Teukolsky equation by writing
\begin{equation}
    \Psi_4=\sum_{\ell,m}e^{-i\omega_{\ell m}t} e^{i m \phi} \psi_{\ell m}(r,\theta), \quad \psi_{\ell m}(r,\theta)=R_{\ell m}(r){}_{-2}S_{\ell m}(\theta),
\end{equation}
where ${}_{-2}S_{\ell m}(\theta)$ are the spin-weighted spheroidal harmonics. The latter solves the angular equation (for generic $s$)
\begin{equation}
\label{eq:angular_teuk}
    \frac{\dd}{\dd z}\left((1-z^2)\frac{\dd }{\dd z} \,{}_{s}S_{\ell m}\right) + \left(a^2\omega^2 z^2-2 a \omega s z +s+ \Lambda + 2 a \omega m- a^2\omega^2-\frac{(m+sz)^2}{1-z^2} \right){}_{s}S_{\ell m}=0,
\end{equation}
with $\Lambda$ being the angular separation constant and we defined $ z=\cos \theta$.
The radial part can be written, also for generic $s$, as
\begin{equation}
    \Delta^{-s+1}\frac{\dd}{\dd r}\left(\Delta^{s+1}\frac{\dd R_{\ell m}}{\dd r}\right) +V(r)R_{\ell m}=0,
    \label{tt1}
\end{equation}
where
\begin{equation}
    V(r)=\left(K^2 -2is(r-M)K +\Delta(4is\omega r - \Lambda)\right),
\end{equation}
\begin{equation}
    K= \omega\left(r^2+a^2\right)-am.
\end{equation}
To simplify the equation we can substitute
\begin{equation}
    R_{\ell m}(r)= \frac{\Delta^{-s}}{\sqrt{r^2+a^2}}\, u(r),
\end{equation}
and employ the tortoise coordinate $r_*$, defined as 
\begin{equation}
    \frac{\dd r}{\dd r_*}= f(r) = \frac{\Delta}{r^2+a^2}.
\end{equation}
After extracting an overall factor $\Delta^s/(r^2+a^2)^{3/2}$, Eq. (\ref{tt1}) is written as 
\begin{equation}
    u''(r_*) - \frac{2s (r-M)}{a^2+r^2} u'(r_*) + \left[\frac{V(r)}{(a^2+r^2)^2}- \frac{g(r)\Delta}{(a^2+r^2)^4} \right]u(r_*)=0,
\end{equation}
where
\begin{equation}
    g(r)=a^4 (2 s+1)+a^2 r (2 M (s-2)+2 r s+r)+2 M r^3 (s+1).
\end{equation}
We now expand for large $\ell$, with $m=\ell$, keeping only the leading order.
Notice that $g(r)$ does not depend on $\ell$, and therefore contributes only to subleading orders.
By taking the following approximation for the frequencies
\begin{equation}
    \omega = \frac{\ell}{b_0} + \mathcal{O}(\ell^0),
\end{equation}
the potential turns out to be 
\begin{equation}
    V(r)= \frac{\ell^2}{b_0^2}\left[\left(r^2+a^2- ab_0 \right)^2 -\Delta (b_0-a)^2\right] + \mathcal{O}(\ell),
\end{equation}
giving rise to the following  equation at leading order in $\ell$
\begin{equation}
\label{eq:lightring}
    u''(r_*) + \frac{l^2}{b_0^2}\frac{\left(r^2+a^2- ab_0 \right)^2 -\Delta (b_0-a)^2}{(a^2+r^2)^2}u(r_*)=0.
\end{equation}
We ignored the first derivative because it can be restored by sending 
\begin{equation}
    u\to u \, \exp\left(\int \frac{s (r-M)}{a^2+r^2}\right),
\end{equation}
and this change does not affect the potential at leading order in $\ell$. 
Furthermore, all the dependence with respect to $s$ disappears, in line with the intuition that in the eikonal limit all massless perturbations behave in the same way, following the Hamilton-Jacobi equation.

At the light ring $r=r_0$, we have 
\begin{equation}
    V(r_0), \qquad V'(r_0)=0,
\end{equation}
so that  around this point, Eq. (\ref{eq:lightring}) is written as  
\begin{equation}
\label{eq:lightring_approx}
    u''(r_*) + \frac{1}{2} \frac{f(r_0)^2}{(r_0^2+a^2)^2}V''(r_0) \,\delta r_*^2\, u(r_*)=0,
\end{equation}
where $\delta r_*=r_*-r_{0*}$.
Finally, to match to the Penrose limit, it is necessary to change variable from $\delta r_*$ to $x_1$, whereby using the chain transformation
\begin{equation}
    x_1 = \frac{r_0}{\sqrt{\Delta_0}}\delta r, \quad \delta r = f \delta r_*, \quad \Delta_0=\Delta|_{r_0}
\end{equation}
we find that 
\begin{equation}
    q \,\delta r_* =x_1, \qquad q= \frac{r_0}{\sqrt{\Delta_0}}f.
\end{equation}
In then new variable $x_1$ we have 
\begin{equation}
    \partial_{x_1}^2 u + \frac{1}{2} \frac{f(r_0)^2}{(r_0^2+a^2)^2}\frac{V''(r_0)}{q^4} \,x_1^2\, u=0,
\end{equation}
and the coefficients between the equation in Penrose limit match exactly with the above equation for 
\begin{equation}
    \alpha^2 = \frac{a^2-b_0^2+6r_0^2}{r_0^4}, \quad p_v^2=\frac{\ell^2}{b_0^2}.
\end{equation}
After this step we are now ready to perform the WBK matching between the two solutions. 
We will follow what done in Refs.\cite{Yang_2012,Kehagias:2025ntm}, performing the matching both for the angular and the radial part.

Starting from Eq. (\ref{eq:lightring}) and Eq. (\ref{eq:lightring_approx}) and defining
\begin{align}
\label{eq:eikonal_quantities}
    S_r^{(\ell)}(r) &=\int^r\frac{\sqrt{V(r')}}{\Delta(r')}\dd r',\nonumber\\
    S^{(\ell)}_{\theta}(\theta)&=\ell\int^{\theta}\sqrt{\cos^2\theta'\left(\frac{a^2}{b_0^2}-\frac{1}{\sin^2\theta'}\right)}\dd \theta',
\end{align}
we can write an eikonal solution as 
\begin{equation}
    \psi(r,\theta)\simeq e^{i S_r^{(\ell)}(r)}e^{iS_{\theta}^{(\ell)}(\theta)},
\end{equation}
which are the radial and angular part of the ``principal function'' for the Hamilton-Jacobi equation in Kerr spacetime.
We ignored the Carter constant $\mathscr{C}$ because it acts as a $\mathcal{O}(1/\ell)$ correction in the square root, therefore being subleading.
To approximate this solution around the light ring and connect it with the solution found through the Penrose limit, we must expand the integrals in Eq. (\ref{eq:eikonal_quantities}) around $r\sim r_0$ and $\theta\sim \pi/2$.
We  define now
\begin{equation}
    Q(r)= \frac{l^2}{b_0^2}\frac{\left(r^2+a^2- ab_0\right)^2 +\Delta (b_0-a)^2}{(a^2+r^2)^2},
\end{equation}
such that
\begin{equation}
    S_r(r_*)=\int^{r_*} \sqrt{Q(r_*')}\dd r_*'.
\end{equation}
Close to the light ring, we approximate $Q(r)$ by 
\begin{equation}
    Q\simeq \frac{1}{2}Q_0''\delta r_*^2, \quad \frac{Q_0''}{2q^4}=\alpha^2p_v^2,
\end{equation}
where we have defined
\begin{equation}
Q_0''=\partial_{r_*}^2 Q|_{r0}.
\end{equation}
Notice that we have to take also into account the change of variables from $(r,\theta)$ to $(x_1,x_2)$
\begin{equation}
    \delta \theta = \frac{x_2}{r_0}, \qquad \delta r_*=\frac{x_1}{q}.
\end{equation}
After all the above steps we have finally
\begin{align}
    \psi(r,\theta)&\simeq \exp\left(-\frac{\ell}{2}\sqrt{1-\frac{a^2}{b_0^2}}\,\delta\theta^2\right)\exp\left(\frac{i}{2}\sqrt{\frac{Q_0''}{2}}\delta r_*^2\right) =\exp\left(-\frac{\ell}{2}\sqrt{1-\frac{a^2}{b_0^2}}\,
    \frac{x_2^2}{r_0^2}\right)\exp\left(\frac{i}{2}\sqrt{\frac{Q_0''}{2q^4}}x_1^2\right)\nonumber\\
    &=\exp\left(-\frac{1}{2}\alpha p_v x_2^2\right)\exp\left(\frac{i}{2}\alpha p_vx_1^2\right),
\end{align}
where we have express $\alpha^2$ as 
\begin{equation}
    \alpha^2=\frac{b_0^2-a^2}{r_0^4},
\end{equation}
as can easily be verified. 
Therefore the matching at leading order in $\ell$ of  the solution in the eikonal regime and the solution in the Penrose limit can be reduced in the matching of   only the overall constant $c_{\ell}$ in 
\begin{equation}
    \Psi_4|_{\rm pp}= c_{\ell} e^{-i p_v v -i\bar{p}_u u}e^{-i p_v \alpha(x_1^2-i x_2^2)}\quad\Longleftrightarrow \quad\Psi_4|_{\rm eik} = c_{\ell}e^{-i \omega_{\ell} t}e^{i \ell \phi} e^{i S_r^{(\ell)}(r)}e^{i S_{\theta}^{(\ell)}(\theta)}.
\end{equation}
It can be checked that the frequency at large $\ell$ at leading order is just 
\begin{equation}
    \omega_{\ell}= p_v= \frac{\ell}{b_0}= E,
\end{equation}
where $E$ is the conserved energy of the null geodesic.

Finally the angular part $e^{i S_{\theta}^{(\ell)}(\theta)}e^{il\phi}$ can be projected onto spin-weighted spheroidal harmonics, to extract the overall constant and compare the results with the usual nonlinear ratios found in the literature.
To do so we highlight that, using a WKB approach as in Ref. \cite{Yang_2012}, when $\ell\to \infty$ the inversion points in the angular direction coincide to $\theta_{\pm}=\pi/2$, therefore the solution to project to the spin-weighted spheroidal harmonics is
\begin{align}
    e^{iS_{\theta}^{(\ell)}(\theta)}\to&
\exp\left(-\ell \int_{\pi/2}^{\theta} \dd \theta'\sqrt{\cos^2\theta'\left(\frac{a^2}{b_0^2}-\frac{1}{\sin^2\theta'}\right)}\right)\Theta\left(x-\frac{\pi}{2}\right) \nonumber\\
&+\exp\left( \ell\int^{\pi/2}_{\theta}\dd \theta' \sqrt{\cos^2\theta'\left(\frac{a^2}{b_0^2}-\frac{1}{\sin^2\theta'}\right)} \right)\Theta\left(\frac{\pi}{2}-x\right),
\end{align}
where $\Theta(x)$ is the Heaviside step function.
Finally the spin-weighted spheroidal spherical harmonics at large $\ell$ can be approximated with the scalar spheroidal harmonics, because the dependence with respect to the spin enters at order $\mathcal{O}(1/\ell)$, as it can be noticed from Eq.~(\ref{eq:angular_teuk}). 
We then used the simpler scalar spheroidal harmonics for the plots showing the nonlinear ratio $\mathcal{R}_{\ell\times \ell}$, while we projected onto the spin-weighted spheroidal harmonics for the $\ell=2$ part of the ratio $\mathcal{R}_{2\times \ell}$, albeit in this second case most of our approximations do not apply, partly explaining why the result is not precisely in line with recent works.

\bibliography{draft}
\end{document}